\documentclass[
aps,prd,
nofootinbib,
superscriptaddress,
showpacs,
tightenlines,
]{revtex4}
\usepackage{amsmath}
\usepackage{amssymb}
\usepackage{bm}
\usepackage{color,graphicx}
\usepackage{dsfont}
\begin{document}

\title{Identification of Observables for Quark and Gluon Orbital Angular Momentum} 
%The Empire Strikes Back}

\author{Aurore Courtoy} 
\email{aurore.courtoy@ulg.ac.be}
\affiliation{IFPA, AGO Department, Universit\'e de Li\`ege, B\^at. B5, Sart Tilman B-4000 Li\`ege, Belgium
\\ and nstituto de F'sica, Universidad Nacional Aut—noma de MŽxico,
A.P. 20-364, MŽxico 01000, D.F., MŽxico.}

\author{Gary R.~Goldstein} 
\email{gary.goldstein@tufts.edu}
\affiliation{Department of Physics and Astronomy, Tufts University, Medford, MA 02155 USA.}

\author{J. Osvaldo Gonzalez Hernandez}
\email{jog4m@virginia.edu}
\affiliation{Istituto Nazionale di Fisica Nucleare (INFN) - Sezione di Torino
via P. Giuria, 1, 10125 Torino, ITALY, Italy}

\author{Simonetta Liuti }
\email{sl4y@virginia.edu}
\affiliation{University of Virginia - Physics Department,
382 McCormick Rd., Charlottesville, Virginia 22904 - USA \\ and  INFN, Laboratori Nazionali di Frascati, Via E. Fermi 40, 00044, Frascati RM, Italy.} 

\author{Abha Rajan}
\email{ar5xc@virginia.edu}
\affiliation{University of Virginia - Physics Department,
382 McCormick Rd., Charlottesville, Virginia 22904 - USA}


\begin{abstract}
A new debate  has recently arisen on the subject of orbital angular momentum in QCD, in particular, on its observability, and  on its partonic interpretation. Orbital momentum can be defined in QCD using two different decomposition schemes that yield a kinetic and a canonical definition, respectively.   
We argue that kinetic orbital angular momentum is intrinsically associated with  twist three generalized parton distributions, and it is therefore more readily observable, 
while, due to parity constraints, canonical angular momentum, if defined as suggested in the literature in terms of generalized transverse momentum distributions, cannot be observed in scattering processes involving a single hadronic reaction plane. 
\end{abstract}

\maketitle

\baselineskip 3.0ex
The question of the observability of Orbital Angular Momentum (OAM) in QCD was recently addressed  in Ref.\cite{CGGLR}.
The main thrust of the paper was to identify an observable from DVCS experiments for OAM as given by the second moment of the twist-3 GPD, $G_2$ \cite{Polyakov,Penttinen}. 
Alongside with this identification, some difficulties were pointed out which are inherent with the alternative definition of OAM in terms of Generalized Transverse Momentum Distributions (GTMDs), concerning especially the way these objects would be measured in deeply virtual exclusive scattering processes from the proton. This triggered a series of observations in Ref.\cite{kanazawa}, related to the treatment of parity transformations in Ref.\cite{CGGLR}, some of which we believe are ill-founded. This situation, given the important issues at stake, {\it i.e.} the definition of OAM in QCD, and the possibility of measuring it, necessitates therefore an additional  explanation. 

%Generalized Transverse Momentum Distributions (GTMDs), or the structure functions  parametrizing  the nucleon's unintegrated off forward quark-quark correlation function  \cite{Metz2}. An equivalently important related question which has been simultaneously disputed is   
%%
%%
The recent  discussion follows in the steps of a debate which was  initiated previously (see Refs.\cite{Waka,LeaLor} for reviews)
on the gauge invariant decomposition of the total quark and gluon angular momenta, $J^{q}$, and $J^g$, into 
their respective spin and orbital components. One of the results out of this discussion was that it became clear  that OAM could be defined through a twist three contribution from the relation \cite{Polyakov,HatYos},
\begin{eqnarray}
\label{polyakov:eq}
\int dx \, x \, G_2^q(x,0,0)  =  \frac{1}{2} \left[ - \int dx  x  (H^q(x,0,0) + E^q(x,0,0)) + \int dx \widetilde{H}^q(x,0,0) \right] \rightarrow \int dx \, x \, G_2^q(x,0,0) = -L^q, 
% \frac{1}{2}  \int dx \, x (H(x,0,0) + E(x,0,0))  & = & J^q \\
% \int dx \tilde{H}^q(x,0,0)  & = & \Sigma^q 
\end{eqnarray}
where $G_2^q$ is a specific twist three Generalized Parton Distribution (GPD) appearing in the parametrization of the quark-quark correlation function \cite{Polyakov,Penttinen,Kivel,BKM} ($G_2$ was renamed $\widetilde{E}_{2T}$ in the full classification of GPDs given in Ref.\cite{Metz2}); 
$H^q$, $E^q$, and $\widetilde{H}^q$ are  the twist two GPDs contributing to the observables for Deeply Virtual Compton Scattering (DVCS) processes introduced in \cite{Ji_SR} (see reviews in Refs.\cite{Ji_GPD,BelRad}. 
%Eq.(\ref{polyakov:eq}) re-evaluates and confirms within full QCD the relation originally found in  \cite{Polyakov,Penttinen}. 
$L^q$ is referred to as {\em kinetic} \cite{LeaLor} or {\em mechanical} \cite{Waka} OAM, it appears in the relation \cite{Ji_SR},
\begin{equation}
\frac{1}{2}  \Delta \Sigma + L^q +  J^g = \frac{1}{2},
\end{equation}
and it is at variance with the {\em canonical} OAM, $L_{can}^{q,g}$ which is defined through the decomposition \cite{JM},
\footnote{In our previous paper we used the notation ${\cal L}_{q}= L^q_{can}.$}
\begin{equation}
\frac{1}{2} =  \frac{1}{2} \Delta \Sigma + L_{can}^q + \Delta G + L_{can}^g.
\end{equation}
${L}_q(x)$ and $L_{can}^{q}(x)$ admit the same Wandzura Wilczek (WW) part, $L_q^{WW}(x)$, while they differ in their genuine twist three contribution \cite{Hatta,HatYos},
\begin{subequations}
\label{LWW}
\begin{eqnarray}
L_q(x) & = & L_q^{WW}(x) + \overline{L}_q(x) 
\label{Ldyn}
\\
L^q_{can}(x) & = & L_q^{WW}(x)  + \overline{L}_{can}^{q}(x).
\label{Lcan}
\end{eqnarray}
\end{subequations} 
In the WW limit the two OAM distributions coincide, their differences depend on final state interactions contained in this case in the genuine twist three terms. In particular, $\int dx   \overline{L}_q(x) = 0$, while $\int dx   \overline{L}_{can}^q(x) \neq 0$, in general, so that it contributes to the angular momentum sum rule within this specific decomposition. 

Notwithstanding  the notion that the GPDs that enter  Eq.(\ref{polyakov:eq}) can be observed by measuring  specific DVCS asymmetries and cross sections, to validate this relation it is however  necessary to identify processes where OAM can be observed directly through the twist three GPD, $G_2$. 
This was done in Ref.\cite{CGGLR} where, making use of the expressions from an extensive analysis of  DVCS  at twist three level performed in \cite{Kivel,BKM}, 
we were able to single out  the helicity amplitudes combinations  which contribute to the twist three GPD $G_2$, and to connect this structure function  
 to an observable,  namely the $\sin 2 \phi$ modulation in the longitudinal Target Spin Asymmetry (TSA), $A_{LU}^{\sin 2 \phi}$ \cite{CGGLR}. 
This term has already been measured, and found to be quite substantial at HERMES \cite{HERMES} and CLAS~\cite{Chen:2006na}. It is also presently been analyzed at Jefferson Lab \cite{pisano_avakian}. 

On the other side, canonical OAM  is constructed by parametrizing the unintegrated correlation function \cite{JM} in the following way  \cite{Hatta,LorPas,Yuan,BurkardtF14},  
\begin{equation}
\label{lz_F14}
L^q_{can} = \langle p , \Lambda \mid \int d x^- d {\bf x}_\perp \, i \psi^\dagger ({\bf x} \times {\bf \nabla} )^3 \psi \mid p, \Lambda \rangle 
= -\int dx d^2 k_T \frac{k_T^2}{M^2} F_{14}(x,0,k_T^2,0,0),
\end{equation} 
$F_{14}$ is a specific  GTMD  -- or an unintegrated over intrinsic-$k_T$ GPDs -- appearing in the decomposition of the vector component of the unintegrated quark-quark correlation function at twist two \cite{Metz2}
\begin{eqnarray}
\label{vector}
& W^{\gamma^+}_{\Lambda \Lambda'} & \! \! \! (x,{\bf k}_T,\xi=0,{\bf \Delta}_T; \eta)  =  
%% 1
\frac{1}{2 P^+}  \overline{U}(p',\Lambda') \left[  \gamma^+  F_{11} 
+  \frac{i \sigma^{i+} \Delta_T^i}{2M}  (2 F_{13} - F_{11}) 
 +   \frac{i \sigma^{i+} \bar{k}_T^i}{2M} (2 F_{12} ) +
 \frac{i \sigma^{ij} \bar{k}_T^i \Delta_T^j}{M^2}   \, F_{14}  \right]  U(p,\Lambda) \nonumber \\
%% 2  
& = & \delta_{\Lambda, \Lambda'} F_{11} +  \delta_{\Lambda, -\Lambda'}  \frac{-\Lambda \Delta_1 - i \Delta_2}{2M}  (2 F_{13} - F_{11}) +  
 \delta_{\Lambda,- \Lambda'}  \frac{-\Lambda \bar{k}_1 - i \bar{k}_2}{2M}  (2 F_{12}) + \delta_{\Lambda, \Lambda'} i \Lambda \frac{\bar{k}_1\Delta_2 - \bar{k}_2 \Delta_1}{M^2} F_{14}. 
\end{eqnarray}
The connection of $L^q_{can}$,  with kinetic OAM was discussed in several papers \cite{Waka,LeaLor}.  Indeed Eq.(\ref{lz_F14})  provides a plausible, intuituive identification which is inferred from the definition of canonical OAM originally suggested in \cite{JM}. 
Nevertheless, the fact that one can consider matching OAM onto experimental observables, only through a specific off-forward unintegrated parton distribution, or GTMD, entails various complications, 
from questions on  both its factorizability and renormalizability in QCD, to  the definition of a deeply virtual scattering process which could be sensitive to $F_{14}$.
Such complications are not present for the GPD, $G_2$, although there exists no obvious, straightforward partonic interpretation of this twist three quantity.

In this paper we address once more these issues, with the aim of  amending some of the statements in Ref.\cite{kanazawa} which misrepresent,  and somewhat muddle the important steps towards the observability of OAM  given in our original work \cite{CGGLR}.
Our goal is to provide additional support  for pursuing experiments sensitive to both canonical and kinetic OAM.   
Two of the outstanding questions that we address: {\it i)} the nature of the parton distributions being identified with OAM, including the relative quarks and protons spin configurations involved; {\it ii)} the possibility to observe such configurations.

\vspace{0.3cm}
The message we would like to convey with this paper is that a more profound physical understanding of OAM may emerge only by defining a way to measure it. None of the existing literature indicates or states sufficiently clearly either whether this goal can be met, or which experimental setup it would require.
 
\vspace{0.3cm}

In Ref.\cite{CGGLR} we demonstrated that there was  a fundamental reason behind the claim that it was ``not known how to extract Wigner distributions or GTMDs from experiments" \cite{LorPas},   
namely we explained how this inherent difficulty was a consequence of parity  constraints on the helicity amplitudes which enter the general cross section formulation  
\cite{MuldTang,DieSap}. 

\begin{figure}
\includegraphics[width=9.cm]{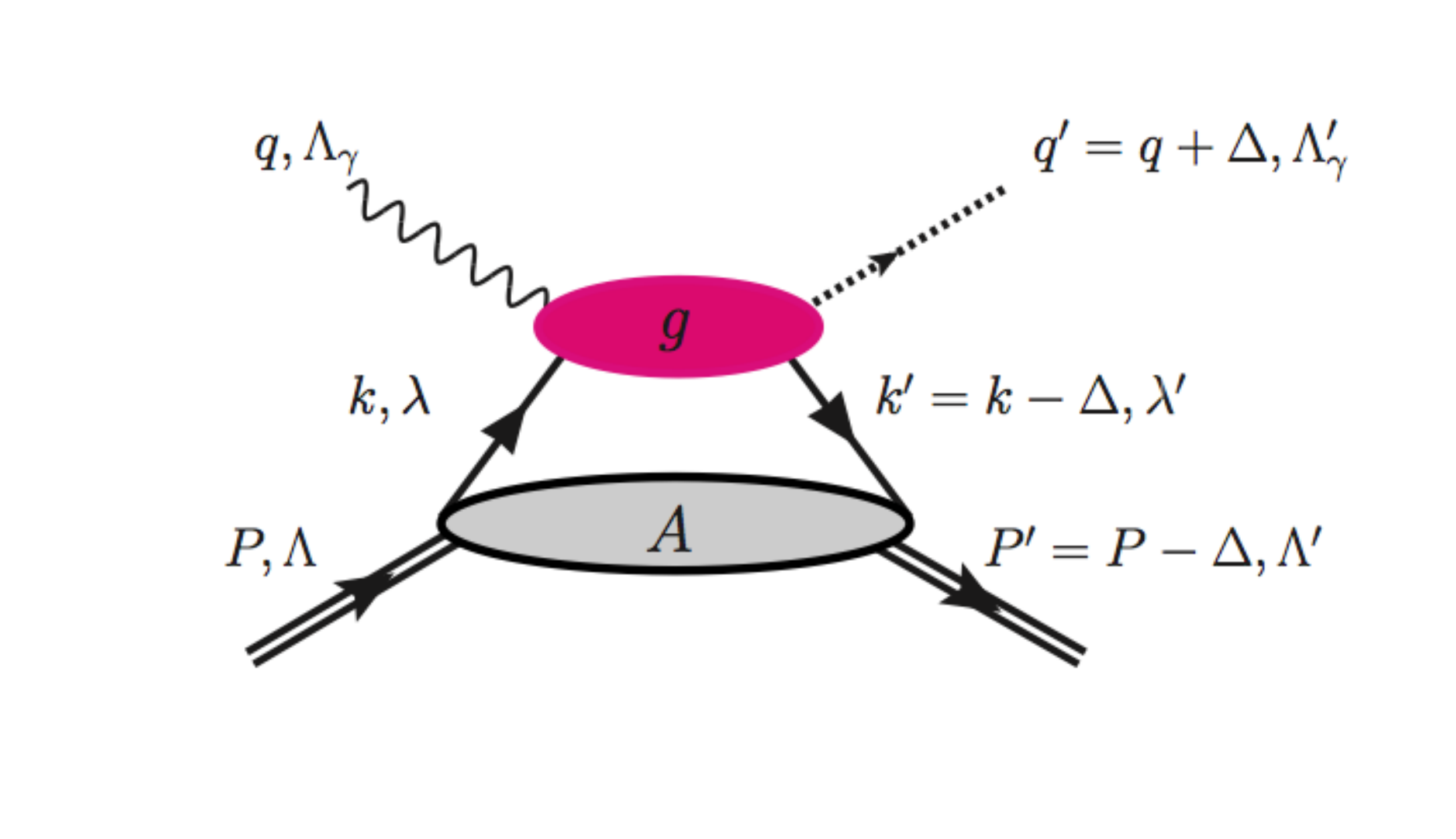}
\caption{The DVCS process (the crossed term diagram is not shown in the figure). The incoming (outgoing) particles momenta and helicities are labelled as follows: $k,
\lambda (k',\lambda')$ for the quarks, $P, \Lambda (P',\Lambda')$ for the protons, and $q,
\Lambda_\gamma (q',\Lambda'_\gamma)$ for the photons.}
\label{fig1}
\end{figure}

Differently from the Transverse Momentum Distributions (TMDs) and the Compton Form Factors (CFFs) which can be extracted  from semi-inclusive and deeply virtual exclusive lepton nucleon scattering, GTMDs cannot be obtained from two body scattering processes.
In Figure \ref{fig1} we illustrate for instance the DVCS process from which GPDs are extracted, namely
\[  \gamma^* p \rightarrow \gamma p' \rightarrow  (\gamma^* q \rightarrow \gamma q') \otimes  (q p \rightarrow q' p') \] 
which factors into $\gamma^*$-quark elastic scattering and two body quark-proton scattering (a similar illustration can be given for the TMD case). 
In such a process, it is always possible to define a Center-of-Mass (CoM) system where the two transverse momenta, ${\bf k}_T$ and ${\bf \Delta}_T$ 
cannot be independent from one another ({\it i.e.} they belong to a single hadronic scattering plane).
%As a consequence, the helicity amplitudes structure 
%%%%
%%%% PARITY SECTION
%%%%
%In this Section we argue about the identification of OAM with $F_{14}$ being dubious because of the manifest parity violation of this term, despite it has been recently shown that it conserves Light Front (LF) parity.
%
%%%%
%%%
%%% FIGURE 1
%%%
To extract $F_{14}$ from experiment one first writes the helicity amplitudes for the $\gamma^* p$ scattering process. The quark-proton scattering 
helicity amplitudes content of $F_{14}$ was identified as \cite{CGGLR},
\begin{eqnarray}
\label{F14}
  i  \frac{\bar{k}_T^1\Delta_2 - \bar{k}_T^2 \Delta_1}{M^2} F_{14} & = & A_{++,++} + A_{+-,+-} - A_{-+,-+} - A_{--,--}, 
\end{eqnarray} 
where $\bar{k}_T=(k_T+k'_T)/2$, and we defined, $A_{\Lambda' \lambda', \Lambda \lambda}$, $\Lambda (\Lambda')$ and $\lambda (\lambda')$ being the proton and quark initial (final) helicities, respectively \cite{Diehl_rev}. 
Following Jacob and Wick \cite{JacobWick} one has, 
\begin{equation}
A_{-\Lambda' -\lambda', -\Lambda -\lambda}=\eta_P (-1)^{\Lambda'-\lambda'-\Lambda+\lambda} A_{\Lambda' \lambda', \Lambda \lambda}^*,
\end{equation}
$\eta_P$ being the phase factor accounting for intrinsic parity and spin. 

This shows explicitly that for the $F_{14}$ contribution to the nucleon matrix elements to be non-zero, at least one pair of the helicity amplitudes must be imaginary, at variance with the other spin conserving structure functions. This circumstance was called ``parity-odd" in Ref.\cite{CGGLR}, borrowing the language from the description of standard two body scattering high energy processes where in the CoM the helicity amplitudes would: 1) be real, and 2) describe a situation where the particles momenta are collinear, thus leading to a zero contribution to Eq.(\ref{F14}), unless a violation of parity conservation occurred as the only possible source of a non zero {\it real} contribution. 

However, for GTMDs, by allowing for relative phases among the amplitudes, the combination that forms $F_{14}$ can indeed be imaginary, as one simultaneously moves away from a collinear description, {\it i.e.} as $\bar{k}$ and $\Delta$ are  let to vary independently from one another. The specific combination of amplitudes giving rise to $F_{14}$ is therefore 
consistent with parity conservation so long as one gives up the idea of the quark proton scattering occurring in a single hadronic plane.
We therefore here acknowledge that it is preferable to use an alternative choice of words to ``parity odd" to describe this rather complicated situation. 

%Yet this cannot appear in the CoM or if all  the momenta all lie in the same plane, that is disposing of a single hadronic scattering plane. 

We further reiterate that this should not be confused with {\it whether the matrix element of the Lorentz structure multiplying $F_{14}$ is parity-even or odd}; parity odd matrix elements obviously occur for many other structure functions which do not display similar pathologies in the CoM, {\it e.g.} the structure function $g_1$ measures helicity which is a pseudoscalar. 

In summary, we have re-explained in more detail  the question addressed  in Section 2 of Ref.\cite{CGGLR}: a combination of purely imaginary helicity amplitudes is clearly zero in the hadronic (or CoM) plane. 
%This fact, that is the impossibility of observing  $F_{14}$ by defining a single hadronic plane, in turn points at  the necessity of going beyond the master equation of Ref.\cite{DieSap} for observing OAM, an issue that will be addressed in an upcoming publication.
In Ref.\cite{CGGLR} 
we raised the issue that  the possibility that  $F_{14}$ is measured, or that it could represent, therefore, a viable observable for OAM,  was hampered by
the parity transformation properties of its nucleon matrix element.
$F_{14}$ is not observable in the general framework defined in Refs.\cite{MuldTang,DieSap} because of parity constraints which limit the amplitudes to be purely imaginary, a configuration which is not physically attainable in a single hadronic scattering plane. This problem can be overcome by going beyond the framework summarized by the master formula of  \cite{MuldTang,DieSap},  an issue that will be addressed in an upcoming publication.

%This does not imply that $F_{14}$ cannot represent OAM. 
%In fact, by observing that,
%\begin{equation} 
%\label{sigma}   
%i \sigma ^{ij} \bar{k}_{T \, i} \Delta_{T \, j}   = \epsilon^{ijk} \Sigma_k \bar{k}_{T \, i} \Delta_{T \, j} = {\bf \Sigma} \cdot ({\bf \bar{k}}_T  \times {\bf \Delta}_T) \equiv \Sigma_3 
% ({\bf \bar{k}}_T  \times {\bf \Delta}_T)_3  ,
%\end{equation}
%one sees that  because of the action of $\Sigma$ on the proton matrix elements, one has that $ i \sigma ^{ij} \bar{k}_{T \, i} \Delta_{T \, j} $ {\em transforms in an opposite way 
%to helicity}, namely to the corresponding structures $G_{14}$, in the GTMD sector, and $\widetilde{H}$ in the GPD sector  \cite{CGGLR,Metz2}. 

Differently from helicity, which is promptly observable, the matrix element corresponding to $F_{14}$ is parity even. Although this is the source of the 
measurability issue for $F_{14}$, it  does not interfere with its identification with OAM which is also a parity even quantity. 
In this respect, our work was misrepresented in Ref.\cite{kanazawa}.

%%%%%%%%%%%%%%%%%%%%%%%%%%%%%%%%%%%%%%%%%%%%%%%%%%%%%%%%%%%%%%%%%%%%%%%%%%%%%%%%%
To clear up any confusion, the same result is obtained by using  parity and Light Front (LF)  parity. 
Under LF parity a transverse vector, ${\bf v} \equiv (v_1,v_2)$  undergoes $L(R) \rightarrow R (L)$ (or $v_1 \mp i v_2 \rightarrow v_1\pm  i v_2$), which takes
\begin{equation}
i {\bf \bar{k}}_T \times {\bf \Delta}_T \rightarrow -i {\bf \bar{k}}_T \times {\bf \Delta}_T.
\end{equation}
Under LF parity transformations  the $x$-components change sign but the $y$-components do not. 
%%%%%%%%%%  END Gary changes %%%%%%%%%%%%%%
Both LF parity and LF helicity amplitudes were implemented {\it e.g.} in  chiral quark models while accounting for the zero modes \cite{CRJi}, and more recently in Ref.\cite{LorPas} 
%Lorce, et al. PRD85,114006 (2012) 
where for the Melosh transform, the 3-quark proton state is written with all the helicities that reverse in order  to have the proton helicity reverse.
%
%\footnote{Notice that  helicity, $h= S \cdot {\bf p}$ behaves the same under parity and LF parity: 
%they are the same because in both cases the $z$-component reverses sign.}
%%
In a LF parity transformation, because of the different behavior of LF helicity and 3-momenta, one has,
\begin{eqnarray}
&& \bar{u}(\tilde{p}',  -\Lambda'_{LF})\gamma^0 i\sigma^{ij} \gamma^0\frac{\bar{k}_i \Delta_j}{M^2}u(\tilde{p}, -\Lambda_{LF}) F_{14}(x, 0, {\bf \bar{k}}_T^2, {\bf \bar{k}}_T\cdot{\bf \Delta}_T, {\bf \Delta}_T^2 ; \eta)  \nonumber \\
& =& (\eta_P)^2\, 2i M \frac{\left(\bar{\bf k}_T\times {\bf \Delta}_T\right)_z}{M^2} (\Lambda_{LF} ) \delta_{\Lambda_{LF} \Lambda'_{LF}} \; F_{14}(x, 0, {\bf \bar{k}}_T^2, {\bf \bar{k}}_T\cdot{\bf \Delta}_T, {\bf \Delta}_T^2 ; \eta),
\end{eqnarray}
for which there is a plus sign times the LF helicity, consistently with what we find.

Therefore, the argument  based on which it is important to use LF helicity amplitudes and parity \cite{kanazawa} is inconsequential for the angular momentum decomposition. In fact,  the matrix element defining $L_{can}$, since it transforms opposite to helicity, is parity even in either case. 
%In fact, it should be kept in mind that $F_{14}$ does  not connect directly to the energy momentum tensor derivation given in Ref.\cite{JM}.  
%$F_{14}$ is a term that appears in the parametrization of the unintegrated correlator \cite{Metz2}. 
Again, the additional distinction we make here is that,  notwithstanding the LF parity argument given in \cite{kanazawa}, in order for any observable to be measured with an experimental  setup involving a single hadronic plane, this has to be a parity preserving object.         
So we are not dwelling on the parity argument but concentrating on what is/can be observable.

%On the other hand,  kinetic OAM can be defined from the twist three correlation function which can be measured 
%via the $\sin 2 \phi$ modulation in the longitudinal Target Spin Asymmetry (TSA). This observable was first proposed and discussed in Ref.\cite{CGGLR}.  
%It should be also noticed that the TSA observable can connect in principle to both $L_q^{can}$ and $L_q$, in yet another way defined in Ref.\cite{HatYos} 
%because, as it can be seen from Eqs.(\ref{LWW})  both definitions involve the same WW piece. In $L_q$ the   
%

\vspace{0.3cm}
The recent argumentations  in Ref.\cite{kanazawa},  however, prompted us to  investigate the existence of observables that could be sensitive to $F_{14}$.
 A natural set of reactions that helps us settle our argument of the parity transformation property of $F_{14}$ is to consider neutrino nucleon scattering processes which notoriously admit parity violating terms. 
The study of observables in manifestly parity violating processes  %represented by the hard part 
turns out, in fact, to be  crucial for settling the controversy. 
A complete list of such observables is given in Ref.\cite{Ji_pol} on the deep inelastic ElectroWeak (EW)  structure of the nucleon along with the helicity composition of the various structure functions which are observable in both inclusive unpolarized and  polarized DIS.
Two of the (current conserved) structures appearing at twist two are parity violating namely the structure function $F_3$ in unpolarized scattering, and $A_1$ in longitudinally polarized scattering, while two are parity conserving, $F_1$ in unpolarized scattering, and $G_1$ in longitudinally polarized scattering. 
%$b_ 2$ in transversely polarized scattering is twist 3. 
%Another, more direct way to get at the parity violating $F_{14}$ is to use parity violating neutrino scattering. 
%That replaces the $\gamma^*$ with $W^\pm$ or $Z^0$. 
All four structure functions  appear in the  transversely polarized vector boson scattering process,
$$(W^\pm, \,  Z^0) + {\rm Nucleon} \rightarrow (W^\pm, \, Z^0)' + {\rm Nucleon}^\prime. $$
%$$W^\pm \, {\rm or}\, Z^0 + {\rm longitudinally \, polarized \, Nucleon} \rightarrow (V, \, A) + {\rm Nucleon}^\prime$$
%where the $(V, \, A)$ are vector and/or axial vector weak bosons or hadrons.
Each one corresponds to a specific combination of the helicity amplitudes, $T_{\Lambda_{(W,Z)} \Lambda, \Lambda'_{(W,Z)} \Lambda'}$ where $ \Lambda_{(W,Z)}$ and $\Lambda$ are the $W^\pm, Z$ and proton helicities, respectively. 

We focus first on the  parity conserving process involving $G_1$ which can be written as, %(see Ji, Eqn. 8 for $S_{1,2}$, but now allowing that these functions depend on $(x, \Delta_T, k_T)$, 
\begin{eqnarray}
%-\frac{4\nu}{M^2}(S_1-\frac{Q^2}{\nu}S_2)&=& T_{+1, +; +1, +} - T_{+1, -; +1, -} - T_{-1, +; -1, +} +T_{-1, -; -, -}   \nonumber \\
G_1 & \propto  & T_{++, + +} - T_{+ -, + -} - T_{- +, - +} +T_{- -, - -}   \nonumber \\
& = &  \sum_{\lambda,\, \lambda^\prime}\left( g^{1,1}_{ \lambda \lambda^\prime} \otimes A_{+\, \lambda, \, + \, \lambda^\prime} - g^{1,1}_{\lambda \lambda^\prime} \otimes A_{- \, \lambda,  - \, \lambda^\prime}  
- g^{-1 -1}_{\lambda \lambda^\prime} \otimes A_{+  \lambda, +  \lambda^\prime} + g^{-1-1}_{ \lambda \lambda^\prime} \otimes A_{-  \lambda,  - \lambda^\prime}\right) \nonumber \\
& = &  \sum_{\lambda,\, \lambda^\prime}\left[  g^{1,1}_{ \lambda \lambda^\prime} \otimes (A_{+\, \lambda, + \, \lambda^\prime} - A_{- \, \lambda, \, -  \lambda^\prime}) - g^{-1-1}_{ \lambda \lambda^\prime}  \otimes  (A_{+\, \lambda, \, + \, \lambda^\prime} - A_{- \, \lambda,  - \lambda^\prime})\right]
\end{eqnarray}
where we considered factorization into quark-proton matrix elements, $A_{\Lambda' \lambda^\prime,\Lambda \lambda}$, and the subprocess, 
%Now in the forward limit (and with quark masses =0) 
$$(W^\pm, \,Z^0) + {\rm quark} \rightarrow  (W^\pm, \, Z^0) + {\rm quark}^\prime. $$
%\label{hard}
We see that, 
%% INSERT by Gary
in the near collinear limit with zero mass quarks, the hard process
%in the forward limit, and with quark masses equal to zero, 
is described by the  amplitudes $g^{1,1}_{+ +}$ and $g^{-1 -1}_{- -}$ (Fig. \ref{fig1}).   
%For parity conserving processes these are identical - they are different for the parity violating weak processes. 
With this reduction in amplitudes the sum becomes,
\begin{eqnarray}
\label{g1}
G_1 & \propto & g^{1, 1}_{+ +} \otimes (A_{+,+;+,+} - A_{-,+;-,+}) - g^{-1 -1}_{- -} \otimes (A_{+,-;+,-} - A_{-,-;-,-}) 
\end{eqnarray}
The hard amplitudes in  Eq.~(\ref{g1}) are evaluated by introducing the quarks' electroweak current,
\begin{equation}
J^\mu = \bar{\psi} \gamma^\mu (g_V \mathds{1} - g_A \gamma^5) \psi,
\end{equation}
where $g_V=g_A=1$  in case of charged currents ($W^\pm$ scattering) and,  for neutral current, $g_V$ and $g_A$ represent the weak vector and axial charges of the quarks.\footnote{See, {\it e.g.}, Ref.~\cite{Psaker:2006gj} for notation.}

One finds,
%%%%%%%% GG Changing a bit of this bad notation
\begin{equation}
%g_{\pm \pm} 
g^{\pm 1, \pm 1}_{\pm \pm} 
%=g^{-1, -1}_{\pm \pm} \quad,\nonumber\\
%\nonumber\\
=\frac{q^- \sqrt{2k^+ k'^+}}{\hat {s}} \left[ (g_V^\prime g_V + g^\prime_A g_A) \mp (g^\prime_V g_A + g^\prime_A g_V)\right]\quad,
\end{equation}
where the prime quantities indicate the final state where both the quark and the final boson can be different from the initial ones in off forward scattering. 
%The polarization $\pm 1$ of the boson does not affect the previous relation.
%
Notice that the parity even relation  $g^{11}_ {++} = g^{-1,-1}_{- -}$, is obtained by setting  ``VA" or ``AV" couplings equal to zero, and  therefore it contains only ``VV" or ``AA" couplings, while the  the opposite happens for the  parity odd relation, $g^{11}_ {++} = - g^{-1,-1}_{- -}$. 
One can  write $G_1$ in terms of parity even and odd contributions as,
\begin{eqnarray}
G_1 & \propto & (g_V^\prime g_V + g^\prime_A g_A)\otimes (A_{++,++} - A_{-+,-+} + A_{- -,- -} - A_{+-,+-}) \nonumber \\ 
& - &  (g^\prime_V g_A + g^\prime_A g_V)\otimes (A_{++,++} - A_{-+,-+} - A_{- -,- -} + A_{+ -,+-})
\label{G1}
\end{eqnarray}
where the parity even hard scattering term selects a combination of soft helicity amplitudes that has the structure of  the DIS structure function $g_{1}$ (or equivalently $\tilde{H}$ in the off-forward case, or $G_{14}$ at the GTMD level),  
while the parity odd combination selects the combination that has the structure of $G_{14}$ at the GTMD level,
but that is  zero when integrated over $k_T$ or in the forward limit.% that is for  $W^\pm (Z^0)  + p \rightarrow W^\pm (Z_0) + p$ .
%We can see thatEq.~\ref{F14} gives 0. The same holds for $Z^0$. 

 A similar analysis, for the function $A_1$ gives,
\begin{eqnarray}
A_1 & \propto & (g_V^\prime g_V + g^\prime_A g_A) \otimes (A_{++,++} - A_{-+,-+} - A_{- -,- -} + A_{+ -,+-})  \nonumber \\ 
& - &  (g^\prime_V g_A + g^\prime_A g_V) \otimes (A_{++,++} - A_{-+,-+} + A_{- -,- -} - A_{+-,+-}),
\label{F14_A}
\end{eqnarray}
where the role of the ``VV" or ``AA" couplings has been switched so that the $F_{14}$-like term appears now in the parity even contribution.
Notice that  because the parity violating terms come from the hard process and these select the helicity states by definition, at the correlator level things remain parity even.

In QCD, in the forward limit  \cite{Ji_pol},
\begin{eqnarray}
G_1 &=& \sum_q (g_V g'_V + g_A g'_A) g_1^q + \; {\rm antiquarks} \\
A_1 &= &-\sum_q (g_V g'_A + g_A g'_V) g_1^q + \; {\rm antiquarks}
\end{eqnarray}
where the sum is taken over the quarks flavors.  This relation is valid for GPDs and TMDs as well, since $A_{++,++} - A_{-+,-+} - A_{- -,- -} + A_{+-,+-}=0$.

If a parity violating term in the unintegrated and off-forward QCD amplitudes level survived, it would be seen as either a parity violating contribution, $\propto g_V g'_A + g_A g'_V$, to  the unintegrated parity even $G_1$ type observable, or as a parity conserving contribution, $\propto g_V g'_V + g_A g'_A$, to the parity violating $A_1$ structure function.  
%However, for reactions like $W^\pm+p\rightarrow \rho^\pm + p$, for example, Eq.~\ref{F14} will not give 0, unless there is an integration over $k_T$, or one takes the forward limit. 

Summarizing this part, to make our observation on the role of parity transformations more concretely tied to experiment, we proved that the same helicity combinations defining  $F_{14}$ and $G_{11}$ appear among the observables for deeply virtual scattering processes as parity violating contributions to the (integrated) QCD structure functions. 

To measure $F_{14}$ and $G_{11}$  and be consistent with the parity transformation properties in QCD one needs to define therefore an additional hadronic plane. 
%%%%% Aurora's wording 
Because $F_{14}$ has the kinematic factor for a longitudinally polarized target going to an unpolarized quark and spectator, it is clear that the hadronization process of the active quark will involve unpolarized functions. Also, as GTMDs depend on the momentum transfer, one has to consider exclusive processes, which rules out ``dihadron" fragmentation functions. What is needed is a final state particle that will force $k_T$ to be fixed, or in some narrow range.

An exclusive process of the type: $\gamma^* + p \rightarrow \gamma + \pi^+ + \pi^- + p^\prime$ will be required. The 4-momenta are set as $q + p = q'+ p_1+ p_2 +p^\prime$. There are 5 invariants,  
%{\bf what is $p_3$??}
%\begin{eqnarray}
$s = (p+q)^2, 
t = (p^\prime - p)^2,
s_{12} = (p_1+p_2)^2,
s_{13} = (p_1 + p^\prime)^2,
t_{1} = (q-p_1)^2$.
%\end{eqnarray}
All other invariants can be written in terms of these. 
%In the CM the final state is like the three-body decay products of a particle of mass$^2 = s$. This can be described by two energies and an angle or some other combinations. 
With this kinematical set of variables one can fix the $k_T$ of the incoming quark, as we will elaborate on in future work. 

\vspace{0.3cm}
Finally, we recall that GTMDs were originally introduced as purely theoretical tools to interpret the possible relations among TMDs and GPDs \cite{Metz2} before   
even considering them as viable means to extract physics information on OAM from experiment in their own merit \cite{LorPas}. 
In this respect, the recent developments in \cite{Polyakov,Hatta,HatYos,LorPas,BurkardtF14,CGGLR} allow us to see how both canonical and kinetic OAM can be represented and compared  at the density level. 
For the density distributions describing quarks' canonical OAM, $L_{can}^{q}$ one has,
\begin{equation}
\label{S_L}
 \bar{u}(p',\Lambda') \frac{i \sigma^{ij} \bar{k}_T^i \Delta_T^j}{M^2} u(p,\Lambda) F_{14}  \propto \langle {\bf S}_L \cdot \bar{\bf k}_T \times {\bf \Delta}_T \rangle,
 \end{equation}
while for the kinetic one, $L_q$,
\begin{equation}
\label{tw3_OAM}
 \overline{u}(p',\Lambda')\frac{i \sigma^{ji} \, \Delta_T^j}{M} u(p,\Lambda)  G_2 \propto  \left\langle {\bf S}_L \times {\bf \Delta}_T \right\rangle  \equiv  \langle \epsilon_T^{ij} {\bf \Delta}_{T j} \rangle
 \end{equation}
Both distributions describe quarks that are displaced from the origin in the transverse plane.  For $F_{14}$, the displacement  is obtained through Fourier transformation of the 
quark-quark correlator components defining  this structure function (Ref. \cite{LorPas} and Eq.(\ref{vector})),
%%
%\footnote{Here and in what follows we use the same symbol for the functions and their Fourier transforms in order to avoid introducing too many symbols (these can be distinguished by their arguments).} 
%%
%%%%
%%%
%%% FIGURE 2
%%%
\begin{figure}
\includegraphics[width=10.cm]{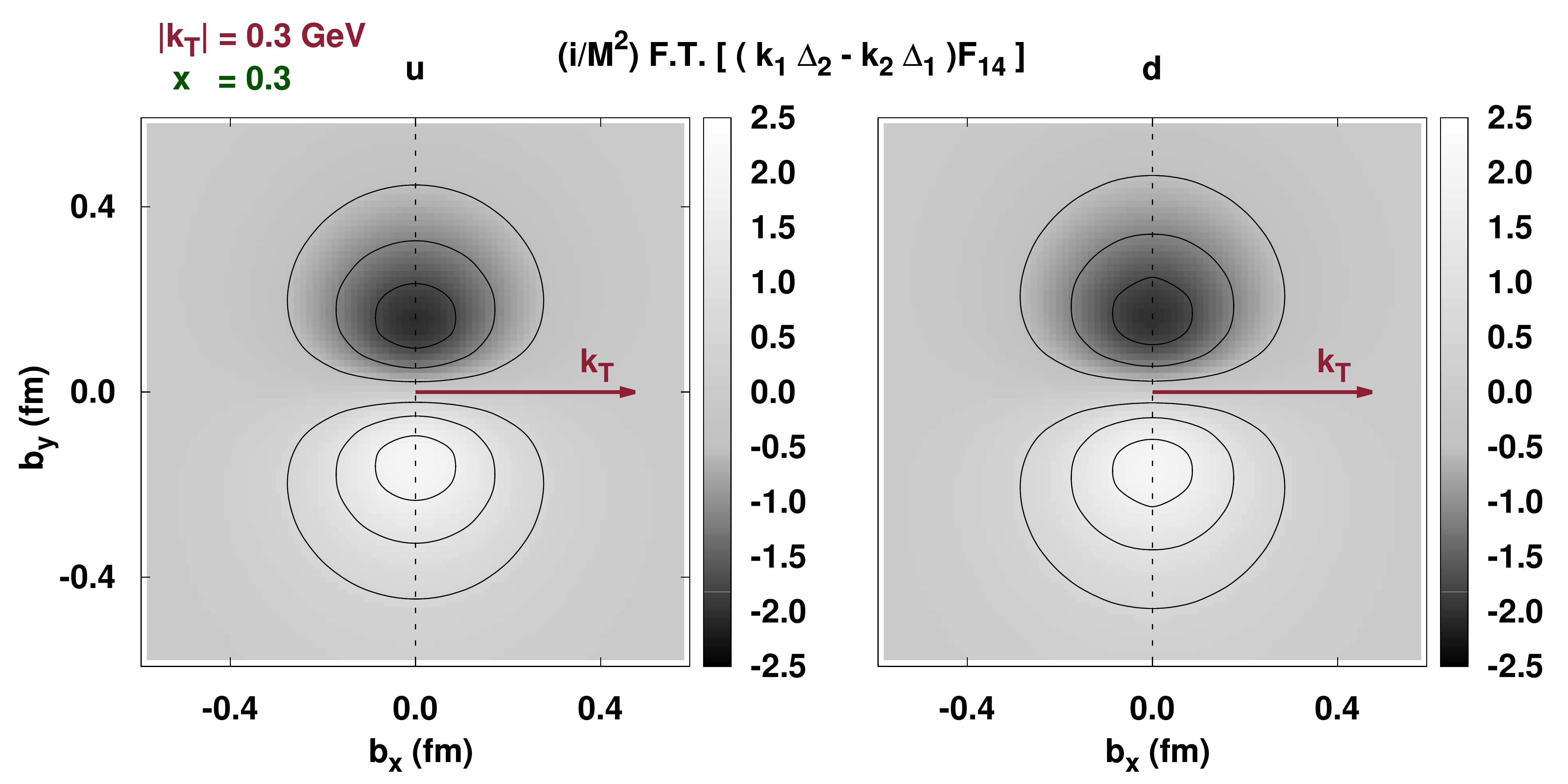}
\caption{Fourier transform of the correlator components defining $F_{14}$, Eq.(\ref{F14:eq}) for $x=0.3$, $k_1=0.3$ GeV, and $k_2=0$. The $u$ and $d$ quarks contributions are represented on the LHS and RHS, respectively.}
\label{fig2}
\end{figure}
\begin{eqnarray}
\label{F14:eq}
{\rm F.T.} \, \left(W^{\gamma^+}_{+ +} - W^{\gamma^+}_{- -} \equiv  i  \frac{\bar{k}_1\Delta_2 - \bar{k}_2 \Delta_1}{M^2} F_{14}(x,0,\bar{\bf k}_T, {\bf \Delta}_T) \right) = -\frac{1}{M^2} \, \epsilon^{ij}_T \bar{k}_T^i \frac{\partial}{\partial b_j} \mathcal{F}_{14}(x,0,\bar{\bf k}_T, {\bf b}),
\end{eqnarray}
where, $\epsilon^{ij}_T=\epsilon^{-+ij}$, and,
\begin{eqnarray}
\label{Fourier}
\mathcal{F}_{14}(x,0,\bar{\bf k}_T, {\bf b}) = \int \frac{d^2 {\Delta_T}}{(2 \pi)^2} \; e^{-i {\bf b} \cdot {\bf \Delta}_T} F_{14}(x,0,\bar{\bf k}_T, {\bf \Delta}_T),
\end{eqnarray}
Notice that $\bar{\bf k}_T$ needs to be kept at a fixed value in order to see this displacement, {\it i.e.} an integration over $\bar{\bf k}_T$ gives a zero result. This configuration corresponds to OAM generated through circular motion in the $x-y$ plane \cite{LorPas}.

For the configuration corresponding to the unintegrated $G_2/\widetilde{E}_{2T}$ (for more details on the specific GTMDs see Refs.\cite{CGGLR,Metz2}),  the distortion in the transverse plane is described by the Fourier transform of the distribution,
\begin{eqnarray}
\label{E2T}
{\rm F.T.} \,  \left(W^{\gamma^i}_{+ +} - W^{\gamma^i}_{- -} \equiv  - i  \epsilon^{ij}_T \frac{\Delta_j}{M}  \,  \widetilde{E}_{2T} (x,0,\bar{\bf k}_T, {\bf \Delta}_T) \right)  =  -\frac{1}{M} \, \epsilon^{ij}_T \frac{\partial}{\partial b_j}  \widetilde{\mathcal{E}}_{2T}(x,0,\bar{\bf k}_T, {\bf b})  \;\;\; (i,j= 1,2),  
%= \int d^2 {\Delta_T} \; e^{-i {\bf b} \cdot {\bf \Delta}}\left [ \left( 2 \widetilde{H}_{2T}(x,0,\Delta_T^2) + E_{2T}(x,0,\Delta_T^2) \right)  + \frac{\Delta_1}{2M}  \widetilde{E}_{2T}(x,0,\Delta_T^2) \right]
\end{eqnarray}
where we used the notation of \cite{Metz2}. In this case ${\bf k}_T$ can be parallel to ${\bf \Delta}_T$, so Eq.(\ref{E2T}) gives a non zero result when integrated over ${\bf k}_T$, and OAM points in the direction orthogonal to both $S_L$ and ${\bf \Delta}_T$, consistently with the representation given in Ref.\cite{JiXiong}. 
Notice that, the distribution functions in principle depend on a vector $n$ that comes from the gauge link \cite{Metz2}. The choice of $n$ is dictated by the leading contribution of the correlator in view of the factorization theorems. That is, the light-cone direction is selected by the hard probe and allows for a separation into hard and soft contributions. The probability density of the unpolarized PDF is valid, as it is well known, only in the light-cone gauge, which masks the dependence on that vector. Once that direction is fixed, the Dirac structure depends on it and the resulting distribution functions will reflect the photon's perspective. We agree that, in such cases, the $z$-direction cannot be changed arbitrarily; an arbitrary rotation of the $z$-axis  was shown to lead to incomplete results in the past~\cite{Courtoy:2008dn}.
However, for  the GTMDs there is no known probe that could disentangle the interplay of the $(x, \xi, {\bf \bar k}_T, {\bf \Delta}_T, {\bf \bar k}_T\cdot{\bf \Delta}_T)$  variables. Hence, the vector $n$ is unconstrained and we believe that it can be freely chosen. For coherence in defining the GTMDs limit to TMDs or GPDs, the authors of Ref.~\cite{Metz2} explcitly chose $n$ to be the light-cone vector $n^{\mu}\equiv (1; 0,0,-1)/\Lambda$ to properly reproduce the path of the appropriate Wilson link.

%%%%%%%%%%%%%%%%%%%%%%%%%%%%%%%%%%%%%%%%%%%%%%%%%%%%%%%%%%%%%%%%%%%%%%%%%%%%%%%%%%%%%%%%%%%%%
%%%%
%%%
%%% FIGURE 3
%%%
\begin{figure}
\includegraphics[width=9.cm]{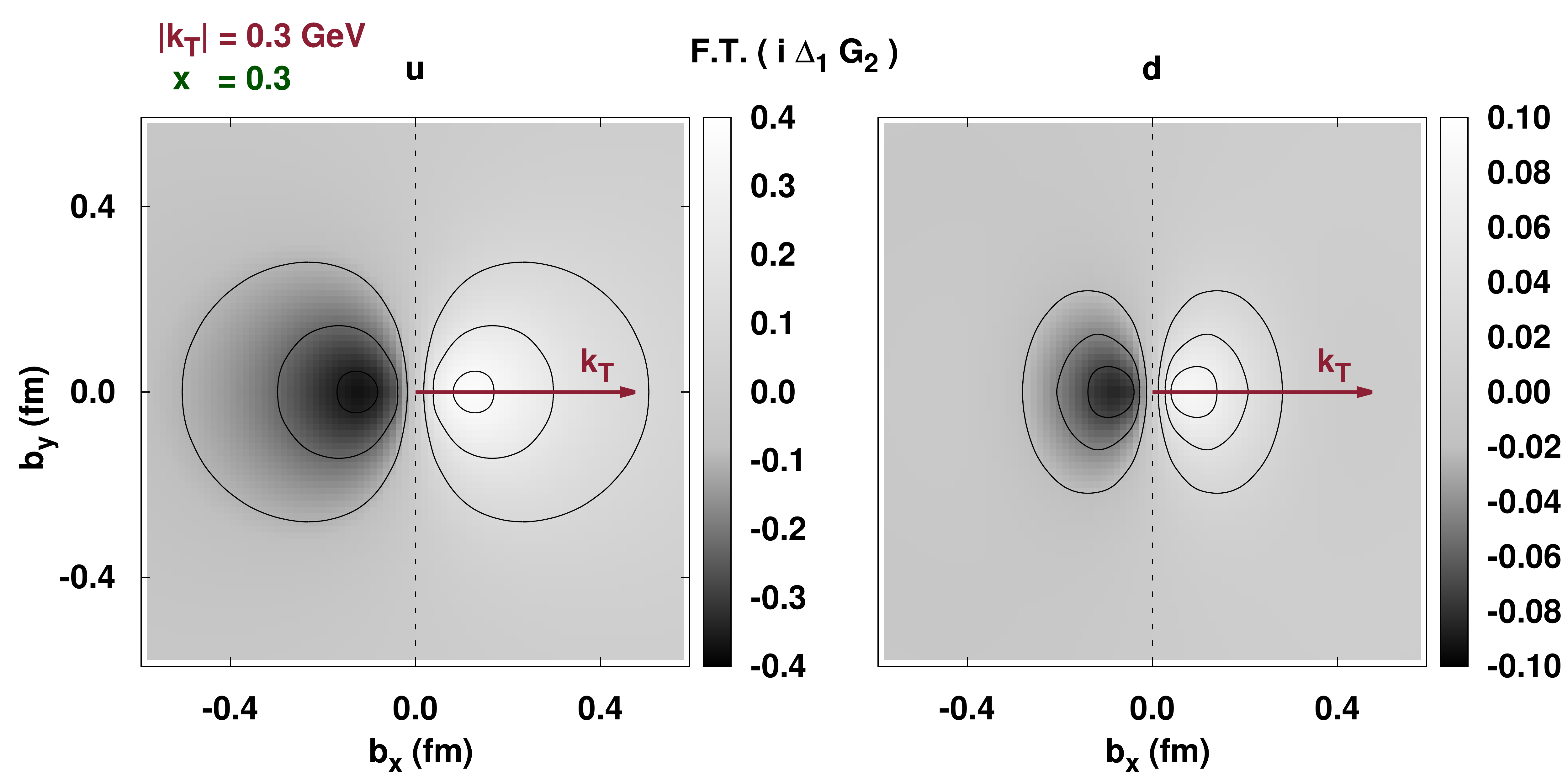}
\includegraphics[width=9.cm]{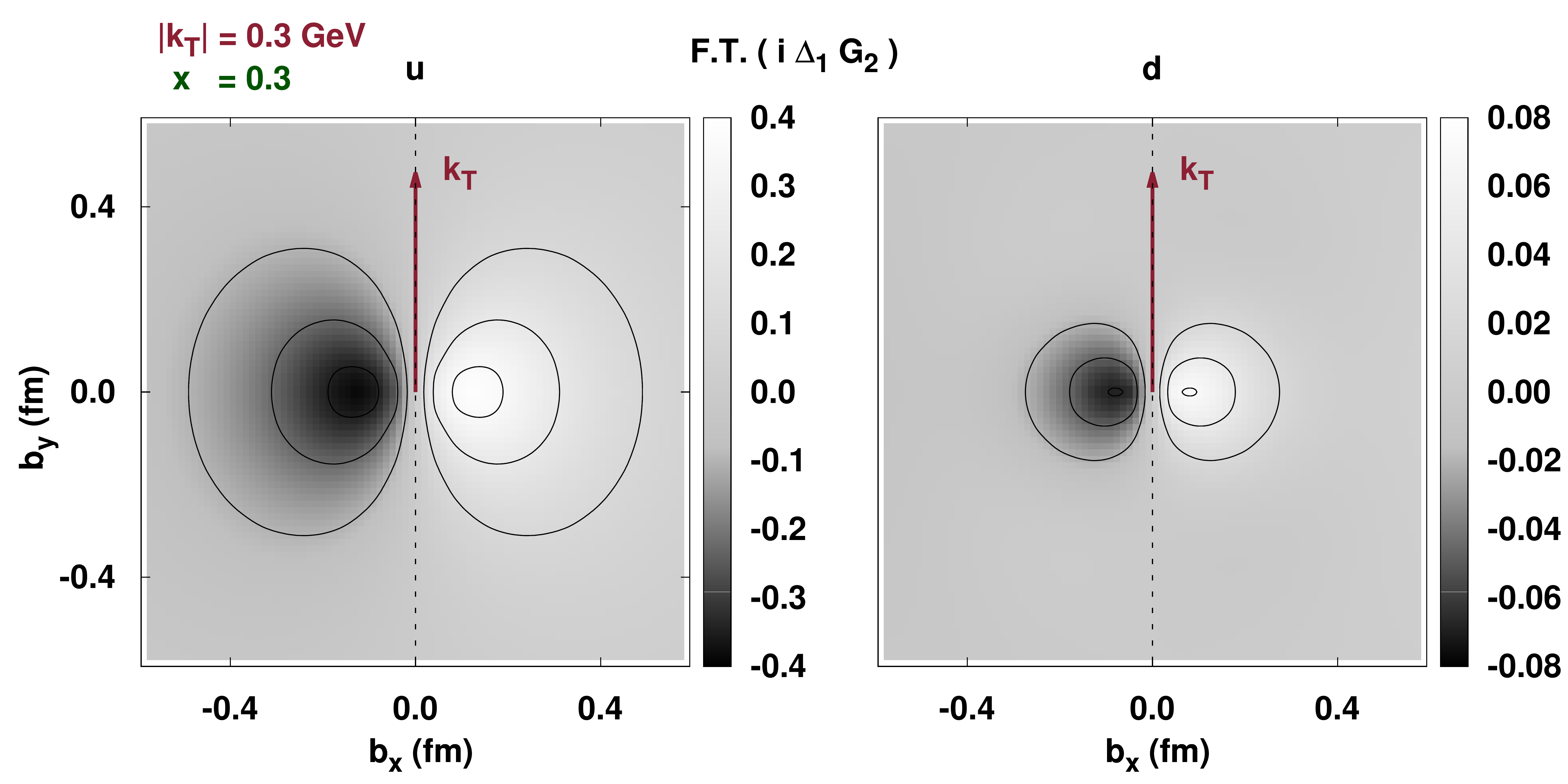}
\caption{Fourier transform of the correlator components defining $\widetilde{E}_{2T}$, Eq.(\ref{E2T}) at $x=0.3$, for the components: $k_1=0.3$ GeV, and $k_2=0$ (upper panels); $k_1=0$, and $k_2=0.3$ GeV (lower panels). For both the upper and lower panels, the $u$ and $d$ quarks contributions are represented on the LHS and RHS, respectively.}
\label{fig3}
\end{figure}

\begin{figure}
\includegraphics[width=10.cm]{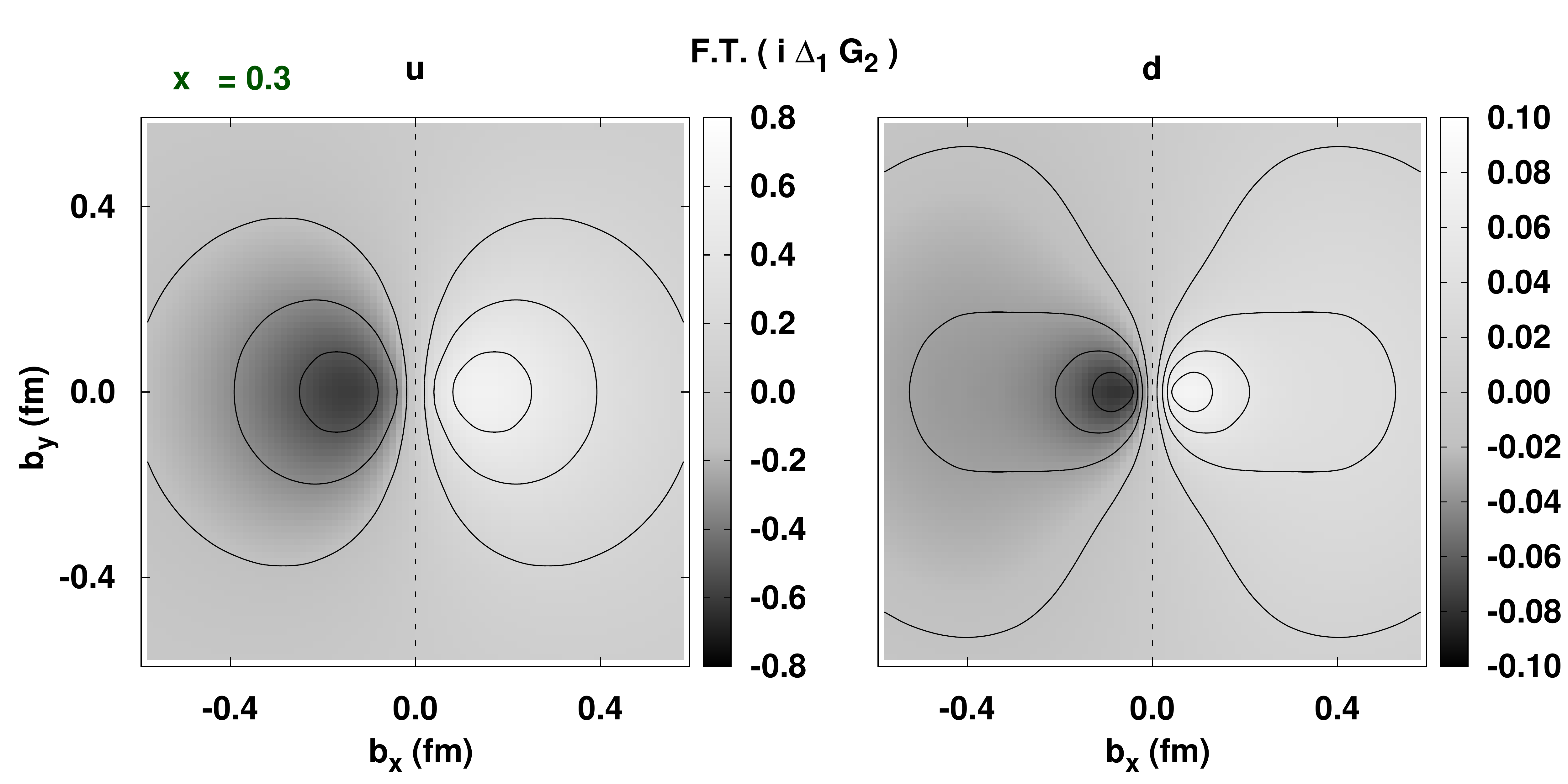}
\caption{Fourier transform of the $u$-quark (L) and $d$-quark (R) correlator components defining $\widetilde{E}_{2T}$, integrated over ${\bf k}_T$, at $x=0.3$.}
\label{fig4}
\end{figure}

In Figure \ref{fig2} and Figure \ref{fig3} we show the distributions in the transverse plane corresponding to Eq.(\ref{F14:eq}) and to the $i=2$ component of Eq.(\ref{E2T}), respectively.  
All functions were obtained in the reggeized quark-diquark picture of Ref.\cite{GGL1,GGLK}.
In Figure \ref{fig4} we show the integrated over ${\bf k}_T$ distribution, for the same value, $x=0.3$ as in Fig.\ref{fig3}. A similar plot could not be drawn for $F_{14}$, since this integrates to zero, thus indicating that these two quantities, although they both represent OAM, correspond to profoundly different partonic configurations.  
%%%%
%%%
%%% FIGURE 4
%%%
%%

%%%%%%%%%%%%%%%%%%%%%%%%%%%%%%%%%%%%%%%%%%%%%%%%%%%%%%%%%%%%%%%%%%%%%%%%%%%%%%%%%%%%%%%%%%

%%%%%% CONCLUSIONS
\vspace{0.5cm}
In conclusion, we have analyzed the issue of observability of both canonical and kinetic OAM.  
Canonical OAM can be identified  with the second moment in $k_T$ of the GTMD $F_{14}$, a formal proof 
that such quantity is related to OAM having been given in Ref.\cite{Hatta}. 
The observability of OAM is however hampered by the fact that $F_{14}$, and an analogously a GTMD in the axial-vector sector, $G_{11}$, cannot be connected to any of the GPDs and/or TMDs, thus making it challenging to define physical processes which would be sensitive to these quantities.
In spite of the fact that non-zero results for $F_{14}$ can be obtained from either models, or by direct calculations on the lattice, a process that selects this quantity has not been yet identified. The physical content of the models, whether these are ``perturbative" or arise from "effective field theories" cannot be taken as a proof  of the existence of an observable. 
%Because the existence of a non-zero result for $F_{14}$ requires a {\em dynamical} imaginary part, although the models are taken at leading order,  there is a confining mechanism that implicitly imposes an imaginary part on the relevant amplitudes.
%which, in turn, allows parity odd components to the helicity structure of the nucleon matrix elements.  
%The assertion that some of the models are ``perturbative" does not rule out a confining mechanism. Each of these has some remnant of confinement. 
%Although $F_{14}$ can be calculated in models this does not imply that it can be (even indirectly) observable. The underlying reason that we explain in this paper is that 
%There is no demonstration that our statement "These non-zero results . . ." is incorrect.  
However, we notice that each of the models explored so far giving a non-zero result for $F_{14}$, carry some remnant of confinement,    
while only one model calculation that clearly does not have confinement - the quark-target model - (as an ``ensemble of free quarks", with no gluon across the vertices) gives zero for $F_{14}$ (see analogous calculation in Ref.\cite{JafJi} on $g_2$). We take this as an indication that the gauge link structure of $F_{14}$ plays a fundamental role, as already suggested in \cite{BurkardtF14} and that, looking at future studies, its connection with the final state interactions implicitly present in the twist three definition of OAM through $G_2/\widetilde{E}_{2T}$ will give key information on the nature of OAM.  

Finally, we reiterate that in our analysis, while reinforcing the use of the LF,  we give a physical motivation for the fact that $F_{14}$ has not yet been associated to any observable, that goes beyond simply stating the issue \cite{LorPas,Metz2}. Our explanation is founded on the transformation properties of the unintegrated correlator under parity which do not allow for the specific combination of helicity amplitudes generating $F_{14}$ to be observed in any given single hadronic plane. This prompts the derivation of an extension of the ``master formula" used so far to describe both semi-inclusive and exclusive lepton-proton scattering \cite{DieSap}. At the same time we point out that the transformation property of the matrix element associated with $F_{14}$ is a distinct issue that should not be confused with the observability of the quantity through its decomposition in quark-proton helicity amplitudes. 
In this respect, $F_{14}$ is consistent with the transformation under parity of OAM, and we do not contend the opposite.

 \vspace{0.3cm}
We benefitted from discussions with many colleagues: Harut Avakian, Ben Bakker, James Bjorken, Elena Boglione, Stan Brodsky, Matthias Burkardt, Michael Engelhardt, Paul Hoyer, Cheung Ji, Cedric Lorc\'e, Stefano Melis, Andreas Metz, Barbara Pasquini, Silvia Pisano, Maxim Polyakov, Masashi Wakamatsu. This work was funded in part by the Belgian Fund F.R.S.-FNRS via the contract of Charg\'ee de recherches (A.C.), and by U.S. D.O.E. grant DE-FG02-01ER4120 (S.L., A.R.). A.R. is grateful to the Universit\'e de Li\`ege for partial funding.

%%%%%%%%%%%%%%%%%%%%%%%%%%%%%%%%%%%%%%%%%%%%%%%%%%%%%%%%%%%%%%%%%%%%%%%%%%%%%%%%


\begin{thebibliography}{99}

\bibitem{CGGLR} %\cite{Courtoy:2013oaa}
%\bibitem{Courtoy:2013oaa}
  A.~Courtoy, G.~R.~Goldstein, J.~O.~G.~Hernandez, S.~Liuti and A.~Rajan,
  %``On the Observability of the Quark Orbital Angular Momentum Distribution,''
  Phys.\ Lett.\ B {\bf 731} (2014) 141
  [arXiv:1310.5157 [hep-ph]].
  %%CITATION = ARXIV:1310.5157;%%
  %4 citations counted in INSPIRE as of 16 Jun 2014

\bibitem{Polyakov} %\cite{Kiptily:2002nx}
%\bibitem{Kiptily:2002nx}
  D.~V.~Kiptily and M.~V.~Polyakov,
  %``Genuine twist three contributions to the generalized parton distributions from instantons,''
  Eur.\ Phys.\ J.\ C {\bf 37} (2004) 105
  [hep-ph/0212372].
  %%CITATION = HEP-PH/0212372;%%
  %18 citations counted in INSPIRE as of 16 Jun 2014

\bibitem{HatYos}
 %%CITATION = ARXIV:1111.3547;%%
  %41 citations counted in INSPIRE as of 16 Jun 2014
  Y.~Hatta and S.~Yoshida,
  %``Twist analysis of the nucleon spin in QCD,''
  JHEP {\bf 1210} (2012) 080
  [arXiv:1207.5332 [hep-ph]].
  %%CITATION = ARXIV:1207.5332;%%
  %12 citations counted in INSPIRE as of 16 Jun 2014

\bibitem{Penttinen} %\cite{Penttinen:2000dg}
%\bibitem{Penttinen:2000dg}
  M.~Penttinen, M.~V.~Polyakov, A.~G.~Shuvaev and M.~Strikman,
  %``DVCS amplitude in the parton model,''
  Phys.\ Lett.\ B {\bf 491} (2000) 96
  [hep-ph/0006321].
  %%CITATION = HEP-PH/0006321;%%
  %64 citations counted in INSPIRE as of 16 Jun 2014

\bibitem{kanazawa} 
%\cite{Kanazawa:2014nha}
K.~Kanazawa, C.~LorcŽ, A.~Metz, B.~Pasquini and M.~Schlegel,
  %``Twist-2 Generalized TMDs and the Spin/Orbital Structure of the Nucleon,''
  Phys.\ Rev.\ D {\bf 90} (2014) 014028
  [arXiv:1403.5226 [hep-ph]].

\bibitem{Waka}  %\cite{Wakamatsu:2012ve}
%\bibitem{Wakamatsu:2012ve}
%\cite{Wakamatsu:2014zza}
%\bibitem{Wakamatsu:2014zza}
  M.~Wakamatsu,
  %``Is gauge-invariant complete decomposition of the nucleon spin possible?,''
  Int.\ J.\ Mod.\ Phys.\ A {\bf 29} (2014) 1430012
  [arXiv:1402.4193 [hep-ph]]. 
%M.~Wakamatsu,
  %``More on the relation between the two physically inequivalent decompositions of the nucleon spin and momentum,''
  Phys.\ Rev.\ D {\bf 85} (2012) 114039
  [arXiv:1204.2860 [hep-ph]].
  %%CITATION = ARXIV:1204.2860;%%
  %21 citations counted in INSPIRE as of 16 Jun 2014

\bibitem{LeaLor} %\cite{Leader:2013jra}
%\bibitem{Leader:2013jra}
  E.~Leader and C.~Lorce,
  %``The angular momentum controversy: What's it all about and does it matter?,''
  %Submitted to: Phys.Rept.
  [arXiv:1309.4235 [hep-ph]].
  %%CITATION = ARXIV:1309.4235;%%
  %16 citations counted in INSPIRE as of 16 Jun 2014
 
\bibitem{Kivel} %\cite{Kivel:2000fg}
%\bibitem{Kivel:2000fg}
  N.~Kivel, M.~V.~Polyakov and M.~Vanderhaeghen,
  %``DVCS on the nucleon: Study of the twist - three effects,''
  Phys.\ Rev.\ D {\bf 63} (2001) 114014
  [hep-ph/0012136].
  %%CITATION = HEP-PH/0012136;%%
  %104 citations counted in INSPIRE as of 16 Jun 2014

\bibitem{BKM} %\cite{Belitsky:2001ns}
%\bibitem{Belitsky:2001ns}
  A.~V.~Belitsky, D.~Mueller and A.~Kirchner,
  %``Theory of deeply virtual Compton scattering on the nucleon,''
  Nucl.\ Phys.\ B {\bf 629} (2002) 323
  [hep-ph/0112108].
  %%CITATION = HEP-PH/0112108;%%
  %314 citations counted in INSPIRE as of 16 Jun 2014

\bibitem{Metz2} %\cite{Meissner:2009ww}
%\bibitem{Meissner:2009ww}
  S.~Meissner, A.~Metz and M.~Schlegel,
  %``Generalized parton correlation functions for a spin-1/2 hadron,''
  JHEP {\bf 0908} (2009) 056
  [arXiv:0906.5323 [hep-ph]].
  %%CITATION = ARXIV:0906.5323;%%
  %53 citations counted in INSPIRE as of 16 Jun 2014

\bibitem{Ji_SR} %\cite{Ji:1996ek}
%\bibitem{Ji:1996ek}
  X.~-D.~Ji,
  %``Gauge-Invariant Decomposition of Nucleon Spin,''
  Phys.\ Rev.\ Lett.\  {\bf 78} (1997) 610
  [hep-ph/9603249].
  %%CITATION = HEP-PH/9603249;%%
  %1201 citations counted in INSPIRE as of 16 Jun 2014

\bibitem{Ji_GPD} X.~Ji,
  %``Generalized parton distributions,''
  Ann.\ Rev.\ Nucl.\ Part.\ Sci.\  {\bf 54}, 413 (2004).
  
\bibitem{BelRad} %\cite{Belitsky:2005qn}
%\bibitem{Belitsky:2005qn}
  A.~V.~Belitsky and A.~V.~Radyushkin,
  %``Unraveling hadron structure with generalized parton distributions,''
  Phys.\ Rept.\  {\bf 418} (2005) 1
  [hep-ph/0504030].
  %%CITATION = HEP-PH/0504030;%%
  %460 citations counted in INSPIRE as of 16 Jun 2014

\bibitem{JM} %\cite{Jaffe:1989jz}
%\bibitem{Jaffe:1989jz}
  R.~L.~Jaffe and A.~Manohar,
  %``The G(1) Problem: Fact and Fantasy on the Spin of the Proton,''
  Nucl.\ Phys.\ B {\bf 337} (1990) 509.
  %%CITATION = NUPHA,B337,509;%%
  %561 citations counted in INSPIRE as of 16 Jun 2014
  
\bibitem{Hatta}  Y.~Hatta,
  %``Notes on the orbital angular momentum of quarks in the nucleon,''
  Phys.\ Lett.\ B {\bf 708} (2012) 186
  [arXiv:1111.3547 [hep-ph]].

\bibitem{HERMES} %\cite{Airapetian:2010ab}
%\bibitem{Airapetian:2010ab}
  A.~Airapetian {\it et al.}  [HERMES Collaboration],
  %``Exclusive Leptoproduction of Real Photons on a Longitudinally Polarised Hydrogen Target,''
  JHEP {\bf 1006} (2010) 019
  [arXiv:1004.0177 [hep-ex]].
  %%CITATION = ARXIV:1004.0177;%%
  %48 citations counted in INSPIRE as of 16 Jun 2014

%\cite{Chen:2006na}
\bibitem{Chen:2006na}
  S.~Chen {\it et al.}  [CLAS Collaboration],
  %``Measurement of deeply virtual compton scattering with a polarized proton target,''
  Phys.\ Rev.\ Lett.\  {\bf 97} (2006) 072002
  [hep-ex/0605012].
  %%CITATION = HEP-EX/0605012;%%
  %103 citations counted in INSPIRE as of 16 Jun 2014
  
\bibitem{pisano_avakian} S. Pisano and H. Avakian, {\it private communication}.

\bibitem{LorPas}  %\cite{Lorce:2011kd}
%\bibitem{Lorce:2011kd}
  C.~Lorce and B.~Pasquini,
  %``Quark Wigner Distributions and Orbital Angular Momentum,''
  Phys.\ Rev.\ D {\bf 84} (2011) 014015
  [arXiv:1106.0139 [hep-ph]].
  %%CITATION = ARXIV:1106.0139;%%
  %52 citations counted in INSPIRE as of 16 Jun 2014

\bibitem{Yuan} %\cite{Lorce:2011ni}
%\bibitem{Lorce:2011ni}
  C.~Lorce, B.~Pasquini, X.~Xiong and F.~Yuan,
  %``The quark orbital angular momentum from Wigner distributions and light-cone wave functions,''
  Phys.\ Rev.\ D {\bf 85} (2012) 114006
 % [arXiv:1111.4827 [hep-ph]].
  %%CITATION = ARXIV:1111.4827;%%
  %30 citations counted in INSPIRE as of 16 Jun 2014

\bibitem{BurkardtF14}
M.~Burkardt,
  %``Parton Orbital Angular Momentum and Final State Interactions,''
  Phys.\ Rev.\ D {\bf 88}, no. 1, 014014 (2013)

\bibitem{DieSap} M.~Diehl and S.~Sapeta,
  %``On the analysis of lepton scattering on longitudinally or transversely polarized protons,''
  Eur.\ Phys.\ J.\ C {\bf 41}, 515 (2005)
  
\bibitem{MuldTang} P.~J.~Mulders and R.~D.~Tangerman,
  %``The Complete tree level result up to order 1/Q for polarized deep inelastic leptoproduction,''
  Nucl.\ Phys.\ B {\bf 461}, 197 (1996)
  [Erratum-ibid.\ B {\bf 484}, 538 (1997)]
 
\bibitem{Diehl_rev} M.~Diehl, Phys.\ Rept.\  {\bf 388}, 41 (2003).

\bibitem{JacobWick} M.~Jacob and G.~C.~Wick,
  %``On the general theory of collisions for particles with spin,''
  Annals Phys.\  {\bf 7}, 404 (1959)
  [Annals Phys.\  {\bf 281}, 774 (2000)].
      
\bibitem{CRJi} C.~E.~Carlson and C.~R.~Ji,
  %``Angular conditions, relations between Breit and light front frames, and subleading power corrections,''
  Phys.\ Rev.\ D {\bf 67}, 116002 (2003)
 
\bibitem{Ji_pol} %\cite{Ji:1993ey}
%\bibitem{Ji:1993ey}
  X.~-D.~Ji,
  %``The Nucleon structure functions from deep inelastic scattering with electroweak currents,''
  Nucl.\ Phys.\ B {\bf 402} (1993) 217.
  %%CITATION = NUPHA,B402,217;%%
  %58 citations counted in INSPIRE as of 16 Jun 2014
   
%\cite{Psaker:2006gj}
\bibitem{Psaker:2006gj}
  A.~Psaker, W.~Melnitchouk and A.~V.~Radyushkin,
  %``Weak Deeply Virtual Compton Scattering,''
  Phys.\ Rev.\ D {\bf 75} (2007) 054001
  [hep-ph/0612269].
  %%CITATION = HEP-PH/0612269;%%
  %8 citations counted in INSPIRE as of 16 Jun 2014
 %%%%%%%%%%%%%%%%%%%%%%%%%%%%%%%%%%%%%%%%%%%
 
\bibitem{JiXiong} %\cite{Ji:2012sj}
%\bibitem{Ji:2012sj}
  X.~Ji, X.~Xiong and F.~Yuan,
  %``Proton Spin Structure from Measurable Parton Distributions,''
  Phys.\ Rev.\ Lett.\  {\bf 109} (2012) 152005
  [arXiv:1202.2843 [hep-ph]].
  %%CITATION = ARXIV:1202.2843;%%
  %45 citations counted in INSPIRE as of 16 Jun 2014
  
   
%\bibitem{Burk_SSA1}   M.~Burkardt,
  %``Quark correlations and single spin asymmetries,''
 % Phys.\ Rev.\ D {\bf 69} (2004) 057501
 % [hep-ph/0311013].
  %%CITATION = HEP-PH/0311013;%%
  %76 citations counted in INSPIRE as of 20 Jun 2014


%\bibitem{Burk_SSA2}   M.~Burkardt,
  %``Sivers mechanism for gluons,''
%  Phys.\ Rev.\ D {\bf 69} (2004) 091501
%  [hep-ph/0402014].
  %%CITATION = HEP-PH/0402014;%%
  %48 citations counted in INSPIRE as of 20 Jun 2014

%\cite{Courtoy:2008dn}
\bibitem{Courtoy:2008dn}
  A.~Courtoy, S.~Scopetta and V.~Vento,
  %``Model calculations of the Sivers function satisfying the Burkardt Sum Rule,''
  Phys.\ Rev.\ D {\bf 79} (2009) 074001
  [arXiv:0811.1191 [hep-ph]].
  %%CITATION = ARXIV:0811.1191;%%
  %32 citations counted in INSPIRE as of 20 Jun 2014
  
\bibitem{GGL1} G.~R.~Goldstein, J.~O.~Hernandez and S.~Liuti,
  %``Flexible Parametrization of Generalized Parton Distributions from Deeply Virtual Compton Scattering Observables,''
  Phys.\ Rev.\ D {\bf 84}, 034007 (2011)
 
\bibitem{GGLK} J.~O.~Gonzalez-Hernandez, S.~Liuti, G.~R.~Goldstein and K.~Kathuria,
  %``Interpretation of the Flavor Dependence of Nucleon Form Factors in a Generalized Parton Distribution Model,''
  Phys.\ Rev.\ C {\bf 88}, 065206 (2013)

\bibitem{JafJi} R.~L.~Jaffe and X.~D.~Ji,
  %``Studies of the Transverse Spin Dependent Structure Function $g$(2) (X, $Q^2$),''
  Phys.\ Rev.\ D {\bf 43}, 724 (1991).
   
\end{thebibliography}
\end{document}